\documentclass[twocolumn,showpacs,preprintnumbers,amsmath,amssymb]{revtex4}

\usepackage{graphicx}
\usepackage{dcolumn}
\usepackage{bm}
\usepackage[colorlinks]{hyperref}

\begin{document}


\title{Geographical networks stochastically constructed \\
by a self-similar tiling according to population}

\author{Yukio Hayashi}
\email{yhayashi@jaist.ac.jp}
\author{Yasumasa Ono}
\affiliation{
Japan Advanced Institute of Science and Technology,\\
Ishikawa, 923-1292, Japan
}

\date{\today}

\begin{abstract}
In real communication and transportation networks, 
the geographical positions of nodes are very important 
for the efficiency and the tolerance of connectivity.
Considering spatially inhomogeneous positions of nodes 
according to a population, 
we introduce a multi-scale quartered (MSQ) network 
that is stochastically constructed by recursive
subdivision of polygonal faces as a self-similar tiling. 
It has several advantages: 
the robustness of connectivity, 
the bounded short path lengths, 
and the shortest distance routing algorithm in a distributive manner.
Furthermore, we show that 
the MSQ network is more efficient with shorter link lengths 
and more suitable 
with lower load for avoiding traffic congestion 
than other geographical networks
which have various topologies ranging from river 
to scale-free networks.
These results 
will be useful for providing an insight into the future design of 
ad hoc network infrastructures. 
\end{abstract}

\maketitle


\section{INTRODUCTION}
Since the common topological characteristics called {\em small-world}
(SW) and {\em scale-free} (SF) have been revealed 
in social, technological, and biological networks, 
researches on complex networks have attracted much attention in this
decade through the historical progress \cite{Newman06}. 
The topological structure is quite different from the
conventionally assumed regular and random graphs,
and has both desirable and undesirable properties: 
short paths counted by hops between any two nodes and
the fault-tolerance for random node removals \cite{Cohen00,Albert00}
on the one hand 
but the vulnerability
against intentional attacks on hubs \cite{Albert00,Callaway00}
on the other hand.
In the definition of a network, 
spatial positions of nodes and distances of links are usually ignored, 
these simplifications are reasonable 
in some networks such as the World-Wide-Web, citation networks, and 
biological metabolic networks.
However, as real-life infrastructures, 
in communication networks, transportation systems, and the power grid, 
they are crucial factors; 
a node is embedded in a mixing of sparse and dense areas according to
the population densities, and a connection between nodes depends on 
communication efficiency or economical cost.
Thus, a modeling of geographical networks is important to understand 
the fundamental mechanism for generating both 
topological and spatial
properties in realistic communication and transportation systems.

Many methods for geographically constructing complex networks 
have been proposed from the viewpoints of 
the generation mechanism and the optimization.
As a typical generation mechanism, 
a spatially preferential attachment is applied in 
some extensions 
\cite{Brunet02,Manna02,Manna03,Nandi07,Wang09} 
of the Barab\'{a}si-Albert (BA) model \cite{Barabasi99}. 
As typical optimizations 
in the deterministic models
\cite{Qian09,Zhou07,Barthelemy06,Gastner06b}, 
there are several criteria 
for maximizing the traffic under a constraint \cite{Qian09}, 
minimizing a fraction of the distance and node degree 
with an expectation of short hops \cite{Zhou07},
and minimizing a sum of weighted link lengths 
w.r.t the edge betweenness as the throughput \cite{Barthelemy06}
or the forwarding load at nodes \cite{Gastner06b}. 
In these methods, various topologies ranging from a 
river network to a SF network emerge
according to the parameter values.
The river network resembles a proximity graph known 
in computer science, which has connections in 
a particular neighbor relationship between 
nodes embedded in a plane.
In constructing these networks, 
it is usually assumed that 
the positions of nodes are distributed uniformly at random, 
and that a population density or the number of passengers
is ignored in communication or transportation networks 
except in some works \cite{Nandi09,Gastner06b,Barthelemy06}.
However, in real data \cite{Yook02}, 
a population density is mapped to the number of
router nodes on Earth, 
the spatial distribution of nodes
does not follow uniformly distributed 
random positions 
represented by a Poisson point process.
Such a spatially  inhomogeneous distribution of nodes is found 
in air transportation networks \cite{Guimera05}
and in mobile communication networks \cite{Lambiotte08}. 

On the other hand, 
geometric methods have also been proposed 
as another generation mechanism,
in which both SW and SW structures are generated by a recursive growing
rule for the division of a chosen triangle 
\cite{Zhang08,Zhou05,Zhang06,Doye05}
or for the attachment aiming at a chosen edge
\cite{Wang06,Rozenfeld06,Dorogovtsev02} 
in random or hierarchical selection.
The position of a newly added node is basically free as far as the
geometric operations are possible,
and has no relation to a population.
Thus, the spatial structure with geographical constraints on nodes and
links has not been investigating enough.
In particular, considering the effects of a population on a geographical
network is necessary to self-organize a spatial 
distribution of nodes 
that is suitable for socio-economic communication 
and transportation requests.

In this paper, as a possibility, 
we pay attention to 
a combination of complex network science and
computer science (in particular, 
computational geometry and routing algorithm) 
approaches.
This provides a new direction of research on self-organized networks
by taking into account geographical densities of nodes and population.
We consider an evolutionary network
with a spatially inhomogeneous distribution of nodes 
based on a stochastic point process.
Our point process differs from the tessellations for 
a Voronoi partitioning 
with different intensities of points \cite{Blaszczyszy04} 
and for a modeling of crack patterns \cite{Nagel07}.
We aim to develop a future design method of ad hoc networks, 
e.g., on a dynamic environment which 
consists of mobile users, for increasing
communication requests, and wide-area wireless and wired
connections.
More precisely, 
the territory of a node defined as the nearest access point 
is iteratively divided for load balancing of communication requests
which are proportional to a population density in the area. 
A geographical network consisting of a self-similar tiling is
constructed by recursive subdivision of faces 
according to a population.
It is worth noting that 
positions of nodes and a network topology are simultaneously
decided by the point process in a self-organized manner.
Furthermore, 
the geographical network has several advantages \cite{Hayashi09}: 
the robustness of
connectivity, the short path lengths, and the decentralized routing
algorithm \cite{Bose04}.
Taking these advantages into consideration, 
we generalize the point process biased by a population for constructing 
a geographical network, 
and investigate the traffic load on the shortest distance routing.

The organization of this paper is as follows.
In Sec. II, 
we introduce a more general network model for self-similar tilings 
than the previous model \cite{Hayashi09}
based on triangulations.
By applying the geometric divisions, 
we construct a geographical network according to a given population. 
In Sec. III, 
we show the properties of the shortest path 
and the decentralized routing without a global table for 
packet transfers as applied in the Internet.
We numerically investigate the traffic
load in the proposed network, 
comparing it with the load in other geographical networks.
In particular, we show that 
our geographical network is better than the state-of-the-art
geographical networks in terms of shorter paths and link lengths, 
and of lower load 
for avoiding traffic congestion.
In Sec. IV, we summarize the results and briefly discuss further
studies.

\section{MSQ NETWORK MODEL} \label{sec2}
\begin{figure}[htb]
\includegraphics[height=40mm]{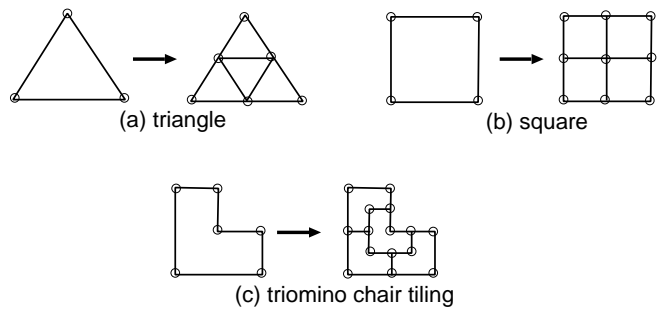} 
\caption{Basic process of the division.}
\label{fig_subdivision}
\end{figure}

\begin{figure}[htb]
\includegraphics[height=70mm]{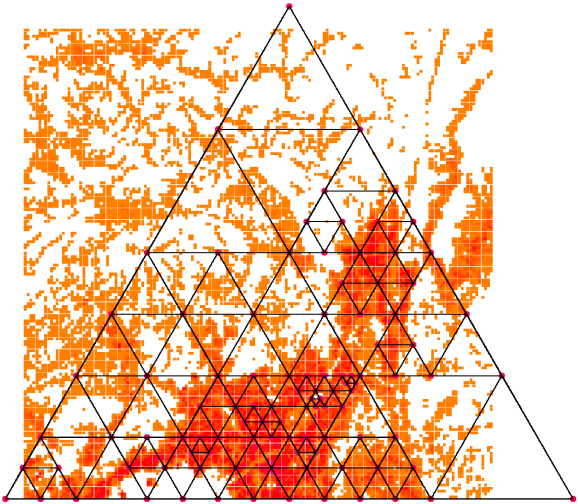} 
\caption{(Color online) Example of the proposed 
geographical network. 
The mesh data of population statistics 
(consisting of 
$8^{2} \times 10^{2} \times 4 = 25600$ blocks for $80 {\rm km}^{2}$)
provided by the Japan Statistical Association
are mapped onto each triangle
by counting the number of people in the space.
From light (orange) to dark (red),
the gradation on a mesh block 
is proportionally assigned to the population.
In the middle of the main island of Japan, 
almost all of the blocks consist of flat lands and mountains, and
do not come in contact with the sea 
except in the left bottom curved white area.}
\label{fig_geo_nets}
\end{figure}

We introduce a multi-scale quartered (MSQ) network, 
which is stochastically constructed by a
self-similar tiling according to a given population. 
Let us consider the basic process of network construction 
\cite{Hayashi09}. 
Each node corresponds to
a base station for transferring packets, 
and a link between two nodes corresponds to a
wireless or wired communication line.
Until a network size $N$ is reached, 
the following process is repeated from an initial configuration which
consists of equilateral triangles or squares. 
At each time step, a triangle (or square) face 
is chosen with a probability 
proportional to the population 
in the space of the triangle (or square). 
Then, as shown in Figs. \ref{fig_subdivision}(a) and 
\ref{fig_subdivision}(b), 
four smaller triangle (or square) faces are created
by adding facility nodes at the intermediate points on 
the communication
links of the chosen triangle (or square). 
This process can be implemented autonomously 
for a division of the area with the increase
of communication requests.
Thus, a planar network is self-organized on a geographical space. 
Figure \ref{fig_geo_nets} shows an example of the geographical MSQ 
network according to real population data.
If we ignore the reality for a distribution of population, 
the MSQ network includes a Sierpinski gasket obtained by
a special selection 
when each triangle, except the central one, is hierarchically divided.

\begin{figure}[htb]
\includegraphics[height=67mm]{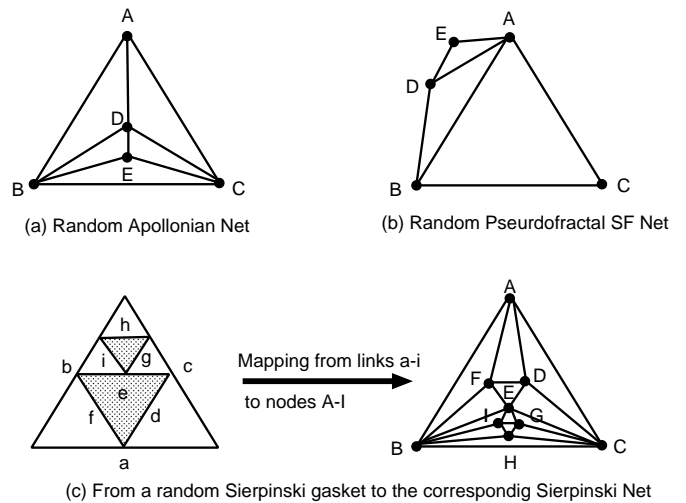} 
\caption{Other geometric networks.
From an initial triangle ABC, the following processes are repeated.
(a) At each time step, a triangle is randomly chosen.
Then, a new node is added (e.g., at the barycenter 
as a well-balanced position) 
and connected to the three nodes of the chosen triangle 
for the division.
(b) Instead of a triangle, an edge is randomly chosen.
Then, a new node is added at an exterior point, 
and connected to the two ends of the chosen edge.
(c) After a random selection of a triangle, 
it is divided into small ones, 
and the center (shaded part) is removed 
to construct a random Sierpinski gasket.
Then, the corresponding random Sierpinski network is obtained from the 
mapping of nodes and contacts to links between them.
Note that the Sierpinski network has a similar structure to the 
Apollonian network rather than the Sierpinski gasket.
}
\label{fig_RAN}
\end{figure}

The state-of-the-art geometric growing network models 
\cite{Zhang08,Wang06,Rozenfeld06,
Dorogovtsev02,Zhou05,Zhang06,Doye05} are summarized 
in Table \ref{table_geometric_nets}.
The basic process for network generation is based on the 
division of a triangle or on the extension of an edge with a 
bypass route as shown in Fig. \ref{fig_RAN}.
For these models, we can also consider a mixing of dense and sparse
areas of nodes by selecting a triangle or an edge 
according to a population in the territory.
Although the geographical position of a new node has not been so far
exactly defined
in the geometric processes, 
it is obviously different from that in the MSQ network.
Moreover, in the Pseudofractal SF network \cite{Wang06}, 
the link length tends to be longer than that in the MSQ network,
because a node is set freely at an exterior point.
Whereas the newly added three nodes approach each other 
in the Sierpinski network \cite{Zhang08}
as the iteration progresses, 
they are degenerated (shrunken) to one node as in the random Apollonian
network \cite{Zhou05,Zhang06}.
Thus, we focus on the Apollonian network constructed by 
a biased selection of a triangle according to a population.
In Sec. III, we compare the
topological and the routing properties in the BA-like and the 
Apollonian networks 
with those in the MSQ network.

The above geometric models 
generate SF networks whose degree distribution follows a
power-law.
It is known that a SF network is extremely vulnerable 
against intentional attacks on hubs \cite{Callaway00,Albert00}, 
in particular 
the tolerance of connectivity is more weakened
by geographical constraints on linking \cite{Hayashi06}. 
However, the MSQ network without high degree nodes 
has a quite different property.
The MSQ network consists of trimodal low degrees: 
$k_{1} = 2$, $k_{2} = 4$, and 
$k_{3} = 6$ for an initial triangle
($k_{2} = 3$, and $k_{3} = 4$ for an initial square) configuration. 
Because of the trimodal low degrees without hubs, 
the robustness against both random failures and intentional attacks
is maintained \cite{Hayashi09} at a similar level 
as the optimal bimodal networks \cite{Tanizawa06} with a
larger maximum degree $k_{2} = O(\sqrt{N})$ 
in a class of multimodal networks, which include 
a SF network at the maximum modality $\rightarrow \infty$
as the worst case for the robustness.

In the MSQ network, 
the construction method defined by recursive subdivision of 
equilateral triangles or squares
can be extensively applied to 
self-similar tilings based on 
polyomino \cite{Keating99}, polyiamond \cite{Solomyak97}, 
and polyform \cite{MathWorld}, 
as shown in Fig. \ref{fig_subdivision}(c).
In the general construction on a polygon, 
some links are removed from the primitive tiling 
which consists of equilateral triangles or squares 
in order to be 
a specially shaped face such as a 
``sphinx'' or ``chair'' at each time step of the subdivision.
It remains to be determined whether or not 
the robustness of connectivity 
is weaker than that in our geographical networks based on 
equilateral triangles or squares.
However, 
the path length becomes larger at least, 
as mentioned in Sec. III A.

\begin{table}[htb] 
\begin{footnotesize} \hspace{-1cm}
\begin{tabular}{c|llll} \hline
Model  & Structure & Add new node(s) & Selection \\ \hline
MSQ    & Trimodal  & On the edges of   & According to \\ 
       & low degrees  & a chosen triangle & a population \\
       &              & (or square)       & \\  \hline
Random & SF, SW, & Mapped to the edges & Random\cite{Zhang08} \\ 
Sierpinski & modular & of a removed triangle &     \\ 
       & & in a Sierpinski gasket &     \\ \hline
Apollonian & SF, SW & Interior of a chosen & Random\cite{Zhou05,Zhang06} or \\
           &        & triangle             & hierarchical \\
           &        &   & deterministic\cite{Zhou05,Doye05} \\ \hline
Pseudofractal & SF, SW & Exterior attached to & 
  Random\cite{Wang06} or \\ 
SF     &            & both ends of an edge & hierarchical \\
           &        & or replacing  & deterministic\cite{Rozenfeld06,Dorogovtsev02}
 \\ 
& & each edge & by two parallel paths & \\ \hline 
\end{tabular}
\end{footnotesize}
\caption{Geometically constructed network models.}
\label{table_geometric_nets}
\end{table}

\section{ROUTING PROPERTIES} \label{sec3}

\subsection{Bounded shortest path}
\begin{figure}[htb]
  \includegraphics[height=18mm]{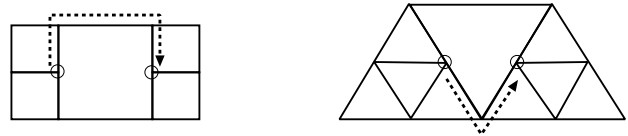} 
\caption{Cases of the maximum stretch factor $t = 2$
for the shortest distance path denoted by a dashed line
and the straight line
directly passing the gap between two nodes marked by circles.} 
\label{fig_stretch}
\end{figure}

The proposed MSQ network becomes 
the {\em $t$-spanner} \cite{Karavelas01},
as a good graph property known in computer science: 
the length of the shortest distance path between any nodes
$u$ and $v$ is bounded by
$t$ times the direct Euclidean distance $d_{uv}$.
A sketch of the proof is shown in the Appendix.
Here, $t$ is called stretch factor which is defined by a ratio of the
path length (a sum of the link lengths on the path) to $d_{uv}$.
Figure \ref{fig_stretch} shows typical cases of 
the maximum stretch factor $t = 2$ in the MSQ network \cite{Hayashi09}. 
When the unit length is defined by an edge of the biggest 
equilateral triangle (or square), the path length 
denoted by a dashed line in Fig. \ref{fig_stretch} is 1 (or 2) 
while the direct Euclidean distance between the two nodes is 1/2 (or 1).

More generally, if we construct a 
network by recursive subdivision
to a self-similar tiling of a polyform \cite{MathWorld}, 
e.g., 
polydrafter: consisting of right triangles, 
polyabolo: consisting of isosceles right triangles, 
or domino: consisting of rectangles, 
then the stretch factor can be greater than $2$ 
[see the U-shaped path in the right of Fig. \ref{fig_subdivision}(c)].
Thus, our network model
based on equilateral triangles or squares
is better for realizing a short routing path 
because of its isotropic property. 
In other geometric graphs, 
the maximum stretch factor becomes larger:
$t = 2 \pi / [3 \cos(\pi / 6)] \approx 2.42$
for Delaunay triangulations \cite{Keil92}, 
and
$t = 2 \alpha \geq 4 \sqrt{3} / 3 \approx 2.3094$
for two-dimensional
triangulations with an aspect ratio of hypotenuse/height
less than $\alpha$ \cite{Kranakis06}, 
whose lower bound is given for the fattest equilateral triangles.
Although
$\Theta$-graphs \cite{Farshi05}
with $K$ non-overlapping cones have
$t = 1 / [\cos(2 \pi / K) - \sin(2 \pi / K)] \rightarrow 1$
asymptotically as $K \rightarrow \infty$,
a large amount of $O(K N)$ links is necessary, and
some links may be crossed.
In general graphs,
even the existence of a bounded stretch factor
is uncertain.
On the other hand, in a SF network, 
the efficient routing \cite{Carmi06} based on the passing through hubs
has a stretch less than 2, which is defined by the ratio of the
number of hops on the routing path to that on the shortest path.
It can be implemented 
by a decentralized algorithm within small memory requirements.

\begin{figure}[htb]
  \includegraphics[height=50mm]{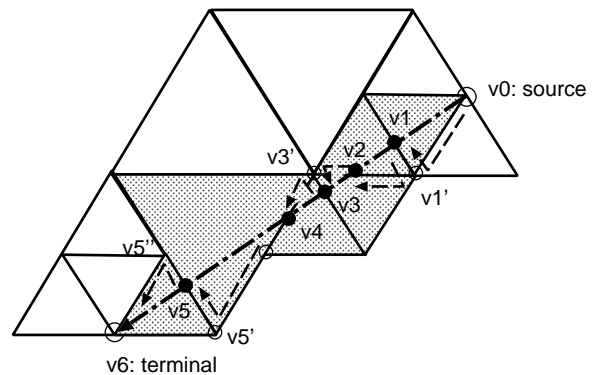} 
\caption{The shortest path obtained by the face routing 
on the edges of the shaded triangles 
that intersect the straight line between the source and the 
terminal nodes.
Two paths of dashed segments 
in the opposite directions (for $v1$-$v1'$ and $v3$-$v3'$) 
are canceled, 
therefore the route 
does not pass the points $v_{1}$ and $v_{3}$, 
and through only the other nodes (marked by open circles)
on the triangles.
Exceptionally, it takes the shortcut $v5'$-$v6$ 
to avoid the detour passing through $v_{5}$ and $v_{5''}$.} 
\label{fig_typical_path}
\end{figure}

Since the geographical MSQ network is planar, 
which is also suitable for avoiding 
the interference among wireless beams, 
we can apply an efficient routing algorithm \cite{Bose04} 
using only local information of the positions of nodes: 
neighbors of a current node, the source, and the terminal 
on a path.
The online version has been developed
in a distributive manner, 
in which necessary information for the routing is
gathered through an exploration within a constant memory. 
As shown in Fig. \ref{fig_typical_path}, 
by using the face routing, 
the shortest distance path can be found on the edges of the faces 
that intersect the straight line between the source and the terminal
nodes.
Note that, in other decentralized routings 
without global information such as a routing table, 
some of them in early work lead to the failure of
guaranteed delivery \cite{Urrutia02}; 
e.g., in the flooding algorithm, multiple redundant
copies of a message are sent and cause network congestion, while
greedy and compass routings may occasionally fall into infinite loops.
On a planar graph, 
the face routing has advantages for the guarantee of a delivery 
and for the efficient search on a short path
without the flooding. 

\subsection{Comparison of $P(k)$ and $P(L_{ij})$ 
with other networks}
We investigate the distributions of degree and of link length 
related to communication costs. 
In the following BA-like networks, 
the positions of nodes are fixed 
as same as in the original MSQ network,
and only the connections are different.
From the set of nodes in the MSQ network, 
a node is randomly selected as the new node 
at each time step in the growing process. 
On the other hand, the positions of nodes in the 
Apollonian network are different from the original ones 
because of the intrinsic geometric construction.

Let us consider 
an extension of the BA model with both effects of 
distance \cite{Brunet02,Manna02,Manna03,Nandi07,Wang09}
and population on linking.
Until a size $N$ is reached, at each time step, 
$m$ links are created from a new node $i$ to 
already existing nodes $j$. 
The attachment probability is given by 
\begin{equation}
  \Pi_{j} \propto d_{ij}^{- \alpha} pop_{j}^{\beta} k_{j}^{\gamma}, 
  \label{eq_BA-like}
\end{equation}
where $\alpha, \beta, \gamma \geq 0$ are parameters 
to control the topology \cite{Nandi07,Zhou07}, 
$k_{j}$ denotes the degree of node $j$, 
and $d_{ij}$ denotes 
the Euclidean distance between nodes $i$ and $j$. 
The newly introduced term 
$pop_{j}$ is not constant but may vary through time. 
If the nearest node from a point on 
the geographical space is changed by adding a new node, 
then the assigned population $pop_{j}$ in the territory of 
the affected node $j$ is updated.
We set the average degree 
$\langle k \rangle = 2 m = 4$ that is the closest integer to 
$\langle k \rangle$ in the MSQ network. 

\begin{figure}[htb]
  \includegraphics[height=80mm]{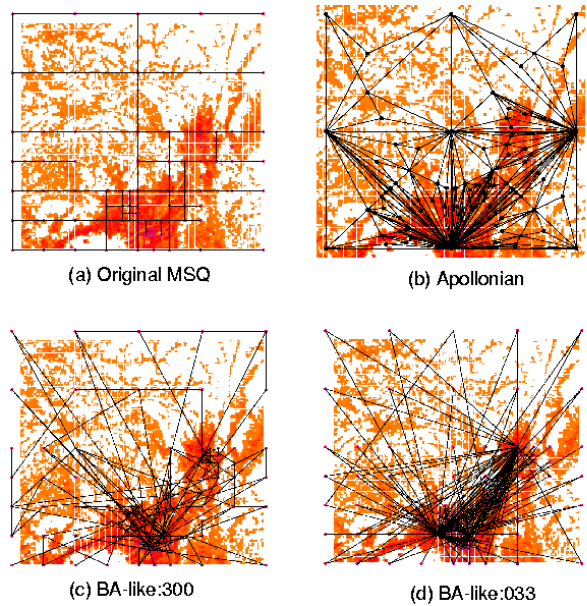} 
\caption{(Color online) Visualizations of the original MSQ, the 
Apollonian, and the typical BA-like networks for $N = 100$. 
As a common feature, many nodes concentrate on 
dark cloudy (red) areas with large populations.
The positions of nodes are same in the MSQ and 
the BA-like networks,
while they are slightly different in the Apollonian network.
(b) Remarkable hubs exist at the center 
and four points on the outer lines.
(c) Almost nearest connections are constructed 
by the attachment of distance. 
(d) Hubs emerge near 
large cities in the dark cloudy (red) areas: 
Kobe, Osaka, and Kyoto 
by the attachments of population and degree.
These results with the nearest connections and hubs 
are similarly obtained in other combinations of 
$\alpha \beta \gamma$ for BA-like networks.
Here, 0 or 3 in the triplet denotes the parameter values for 
$\alpha$, $\beta$, and $\gamma$ in Eq. (\ref{eq_BA-like}).} 
\label{fig_vis_BA-like}
\end{figure}

\begin{figure}[htb]
  \begin{minipage}[htb]{.47\textwidth}
    \includegraphics[height=85mm,angle=-90]{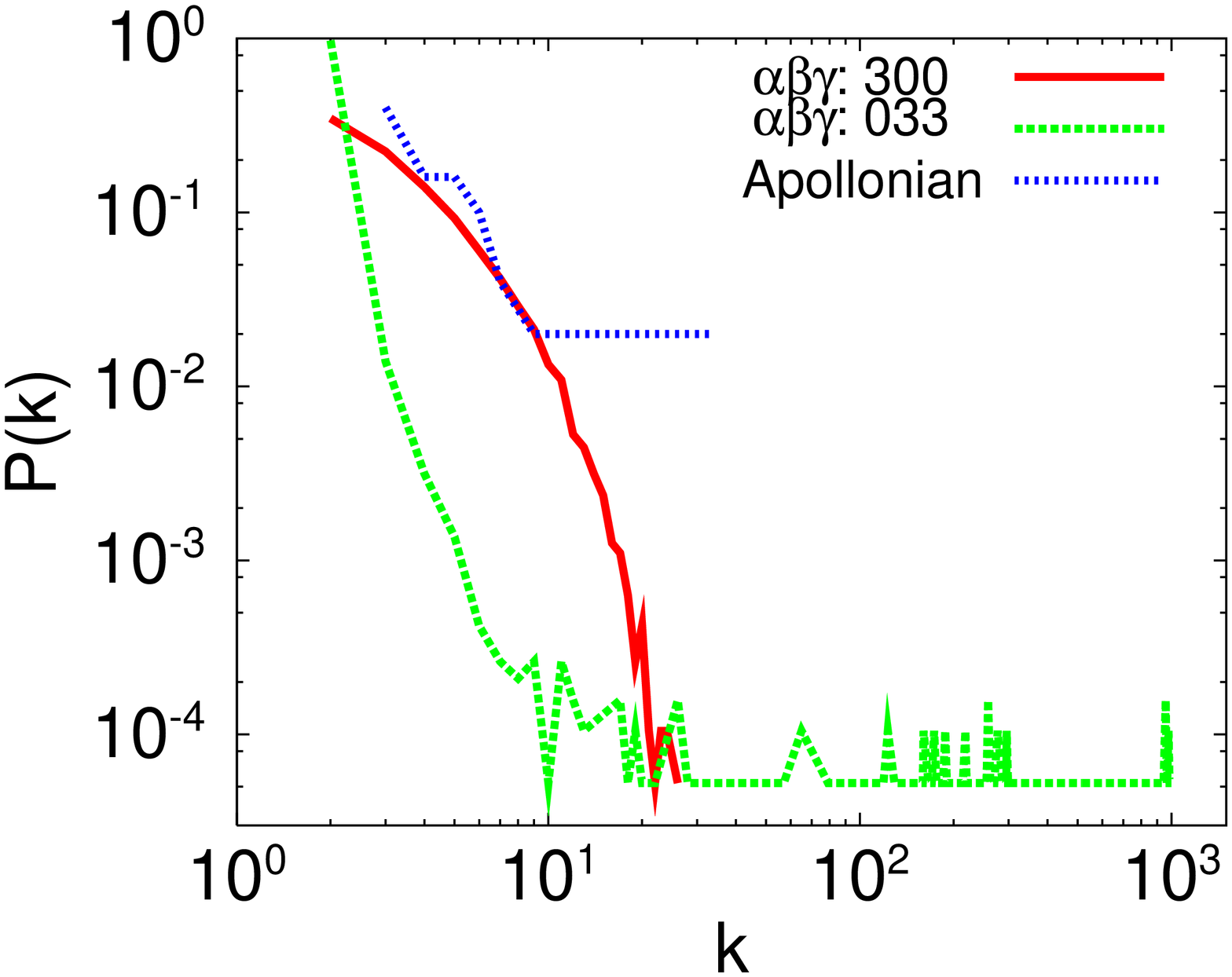} 
    \begin{center} (a) Degree \end{center}
  \end{minipage} 
  \hfill 
  \begin{minipage}[htb]{.47\textwidth}
    \includegraphics[height=85mm,angle=-90]{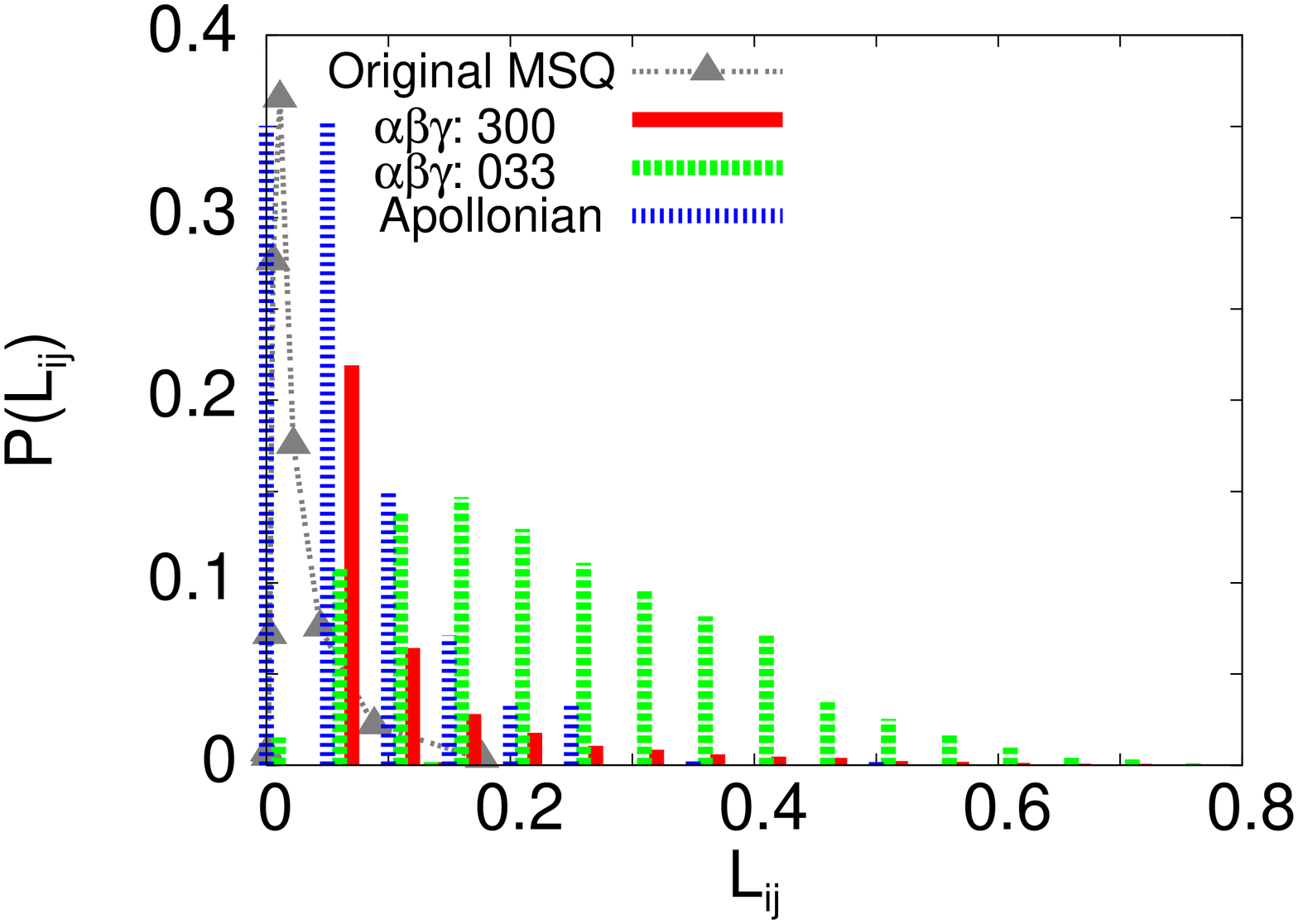} 
    \begin{center} (b) Link length \end{center}
  \end{minipage}
  \caption{(Color online) Distributions of (a) degree 
  and (b) link length 
  in the averaged 50 realizations of the Apollonian and the typical 
  BA-like networks for $N = 1000$. 
  (a) With power-law like behavior, 
  there exists a few huge hubs around $k = 10^{2} \sim 10^{3}$ 
  at the vertical (green) light dashed bars for the BA-like:033 network, 
  and ordinary hubs 
  on the horizontal (blue) dark dotted line for the Apollonian network.
  In spite of the stochastic process, 
  the locations of hubs are fixed e.g., 
  at the center and four points on the outer lines 
  in Fig.\ref{fig_vis_BA-like}(b). 
  Thus, the frequencies are high at some special degrees.
  (b) The histograms are shown with shifts 
  for better discrimination.
  The gray dashed lines with filled triangles show $P(L_{ij})$
  in the original MSQ network.}
  \label{fig_BA-like_Pk_LL}
\end{figure}

Each term in the right-hand side of Eq. (\ref{eq_BA-like}) 
contributes to making a different topology, 
as the value of $\alpha$, $\beta$, or $\gamma$ is
larger in the competitive attachments. 
As shown in Fig. \ref{fig_vis_BA-like}, 
a proximity graph is obtained in the BA-like:300,330 networks
by the effect of distance for $\alpha = 3$, 
and hubs emerge near large cities in the BA-like:030,033,333 networks
by the effect of population for $\beta = 3$, 
while in the BA-like:303,003 networks, 
a few hubs emerge at the positions of randomly selected nodes 
in the early stage of network generation 
without any relation to the population. 
In the following discussion, we focus on BA-like:300,033 networks,
since they are typical with the minimum and the maximum 
degrees or link lengths, respectively, 
in the combination of $\alpha \beta \gamma$ 
except 000,003 
without both effects of distance and population. 
Figure \ref{fig_BA-like_Pk_LL}(a) shows that 
the degree distributions $P(k)$ follow an exponential decay in the 
BA-like:300 network, and a 
power-law like behavior in the BA-like:033
and the Apollonian networks,
which are denoted by a (red) solid line, (green) light dashed line, 
and (blue) dark dotted line, respectively. 
Figure \ref{fig_BA-like_Pk_LL}(b)
shows the distributions $P(L_{ij})$
of link length counted as histograms
in the interval 0.05.
The $L_{ij}$ is normalized by the maximum length 
on the outer square or triangle. 
Some long-range links are remarkable in the BA-like networks, 
while they are rare in the quickly decaying distribution
in the MSQ network.
The average lengths shown in Table \ref{table_ave_link_len}
are around $0.2 \sim 0.3$ in many cases of the BA-like networks, 
and $0.02 \sim 0.14$ in the Apollonian network.
The original MSQ network has 
the smallest link length in less than one digit. 
As the size $N$ increases, 
the link lengths tend to be shorter 
due to a finer subdivision in all of the networks. 
In any case, the BA-like networks have longer links 
even with neighboring connections 
than the corresponding MSQ network, 
while the Apollonian network shows the intermediate result.
Thus, the MSQ network is better than the Apollonian and 
the BA-like networks in 
term of the link length related to a communication cost.

\begin{table}[htb]
\begin{center}
\begin{footnotesize}
\begin{tabular}{c|ccc|c} \hline
BA-like: & & $\langle L_{ij} \rangle$ & & Dominant \\ 
$\alpha\beta\gamma$ & $N = 10^{2}$ & $10^{3}$ & $10^{4}$ & factor \\ \hline
000 & 0.3018 & 0.2380 & 0.1909 & Rand. attach. \\
003 & 0.2757 & 0.2736 & 0.1700 & Degree \\
030 & 0.2759 & 0.2651 & 0.2318 & Population \\
{\bf 033} & {\bf 0.2685} & {\bf 0.2650} & {\bf 0.2622} & Pop. \& deg. \\
{\bf 300} & {\bf 0.1268} & {\bf 0.0436} & {\bf 0.0213} & Distance \\
303 & 0.2089 & 0.1409 & 0.1127 & Dist. \& deg. \\
330 & 0.1789 & 0.0652 & 0.0338 & Dist. \& pop. \\
333 & 0.2318 & 0.2064 & 0.1613 & All of them \\ \hline
Apollonian & 0.1401 & 0.0627 & 0.0184 & \\ \hline
MSQ & 0.0628 & 0.0066 & 0.0033 & \\ \hline
\end{tabular}
\end{footnotesize}
\end{center}
\caption{Average link length $\langle L_{ij} \rangle$
in the original MSQ, the Apollonian, and the BA-like networks.
The $\langle L_{ij} \rangle$ becomes longer in the order: 
the MSQ, the BA-like:300 or the Apollonian, 
and the BA-like:033 networks.
This order is consistent 
with the positions of peaks and the widths of $P(L_{ij})$ in 
Fig. \ref{fig_BA-like_Pk_LL}(b).
Note that 
BA-like:000,003 networks are not geographical, but exceptional, 
because they have no effects of distance and population on linking.
Here, the triplet of 0 or 3 in the first column denotes the values of 
$\alpha$, $\beta$, and $\gamma$.} 
\label{table_ave_link_len}
\end{table}

\subsection{Heavy-loaded nodes and links}

In a realistic situation,
packets are usually more often generated and received at a node, 
as the corresponding population is larger 
in the territory of the node.
Consequently, the spatial
 distribution of the source or the terminal node
is not uniformly random.
Thus, we demonstrate how the traffic load is localized in the case 
when the number of generated and received packets at a node 
varies depending on a population assigned to the node. 

The traffic loads at a node $i$ and through a link $l$
are measured by the effective 
betweenness centralities $B_{i}$ and $\bar{B}_{l}$,
which are defined as follows \cite{Guimera02}:
\begin{equation}
  B_{i} \stackrel{\rm def}{=} \frac{2}{(N-1)(N-2)}
         \sum_{k < j} 
         \frac{b_{k}^{j}(i)}{b_{k}^{j}}, \label{eq_def_bci}
\end{equation}
\begin{equation}
  \bar{B}_{l} \stackrel{\rm def}{=} \frac{2}{(N-1)(N-2)}
         \sum_{k < j}
         \frac{\bar{b}_{k}^{j}(l)}{b_{k}^{j}}, \label{eq_def_bcl}
\end{equation}
where 
$b_{k}^{j}$ is the number of shortest distance paths between 
the source $k$ and the terminal $j$, 
$b_{k}^{j}(i)$ is the number of the paths passing through
node $i$, and 
$\bar{b}_{k}^{j}(l)$ is the number of the paths passing through
link $l$.
The first terms on the right-hand side
of both Eqs. (\ref{eq_def_bci}) and (\ref{eq_def_bcl})
are normalization factors. 
Although the measured 
node $i$ and link $l$ are usually excluded from the sum 
in the definition of betweenness centralities \cite{Freeman91}, 
we include them tanking into account the processes for the 
generation and the removal of a packet 
in these measures $B_{i}$ and $\bar{B}_{l}$ 
in order to investigate all of the traffic loads.

\begin{figure}[htb]
  \includegraphics[height=45mm]{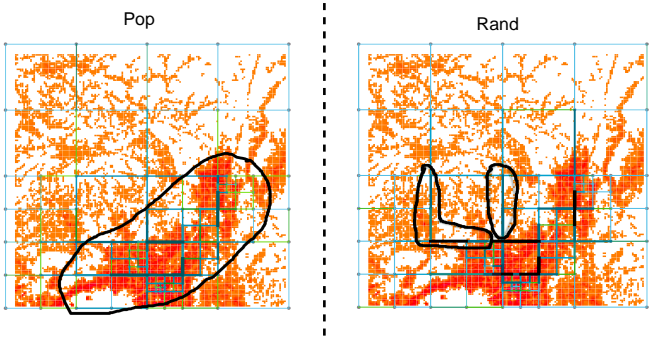}
\caption{(Color online) Visualizations of 
the link betweenness centrality    
$\bar{B}_{l}$ in a MSQ network for $N = 100$. 
Pop (left) and Rand (right) denote the selection patterns of 
a source or a terminal node, which are chosen 
proportionally to the population in the territory
of a node, and uniformly at random, respectively.
From the thin (green) to the thick (blue) vertical or horizontal 
line,
the gradation is proportionally assigned to the value of $\bar{B}_{l}$.
The enclosed bold line in Pop
emphasizes the parts of large $\bar{B}_{l}$ with heavy load, 
which are remarkably shown as thick (blue) links on the paths 
connected to dark cloudy (red) 
areas with large populations in a diagonal direction.
The enclosed bold lines in Rand emphasize
the thick (blue) lines on light gray (orange and white) background 
for underpopulated areas, 
as the other differences between 
the left and the right figures. 
Thus, in Rand, some thick (blue) lines for large $\bar{B}_{l}$
have no relation to the population.
}
\label{fig_MSQ_bc_link}
\end{figure}

\begin{figure}[htb]
  \includegraphics[height=73mm]{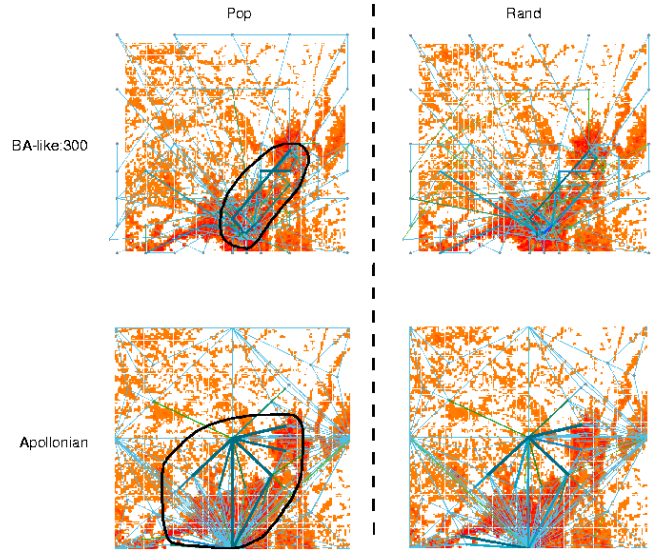}
\caption{(Color online) Visualized examples of 
the link betweenness centrality    
$\bar{B}_{l}$ in the BA-like:300 (top) and the Apollonian (bottom) 
networks for $N = 100$. 
The enclosed bold line in Pop emphasizes that 
the thick (blue) links for large $\bar{B}_{l}$
are on the paths connected to dark cloudy (red) areas 
with large populations.
Pop (left) and Rand (right)
denote the two selection patterns of a source or a
terminal node which is chosen proportionally 
to the population assigned to a node, 
and uniformly at random, respectively.}
\label{fig_BA-like_bc_link}
\end{figure}

Figures \ref{fig_MSQ_bc_link} and \ref{fig_BA-like_bc_link}
show the link load measured by
$\bar{B}_{l}$ for the following 
two selection patterns of packet generations or removals. 
In the first pattern denoted by Pop, 
as the source or the terminal, 
a node is chosen with a probability
proportional to the assigned population to the node (left figures), 
while in the second pattern denoted by Rand, it is chosen 
uniformly at random (right figures).
Therefore, 
the sum in Eqs. (\ref{eq_def_bci}) or (\ref{eq_def_bcl}) has a biased
frequency for each pair of nodes by their populations in Pop, 
while the sum corresponds to the simple 
combination of nodes in Rand. 
In Pop, 
heavy-loaded links denoted by thick (blue) lines 
between the cloudy (red) areas with large populations are observed: 
the routes between well-known cities  
Osaka and Kyoto (in the enclosed bold lines) 
are remarkable with thicker (blue) links 
than in Rand.
Note that the number $N = 100$ of nodes is due
to a comparatively clear appearance of the difference in two patterns. 
At a larger $N$, 
the packet transfer is already biased 
by the spatial concentration of nodes, 
and the assigned population in each territory of node 
is well-balanced by the division process.
Then, the difference of heavy-loaded links
in Rand and Pop tends to disappear.
However, it is worth considering 
such biased selections of packets in a realistic problem setting, 
since the spatially localized 
positions of the heavy-loaded links are not trivially
predictable from the results for the usually assumed Rand.

We further investigate the maximum betweenness 
in the scaling relation $N^{\delta}$ 
which affects traffic congestion \cite{Sreenivasan07}. 
Here, the betweenness is defined by the number of
passings through a link or a node on the shortest distance paths.
In general, a smaller $\delta$ yields a better performance for 
avoiding traffic congestion.
The exponent value $\delta$ depends on both a 
network topology and a routing scheme. 
In this paper, assuming the shortest distance paths obtained by 
the face routing in Sec.\ref{sec3}A, 
we focus on an effect of the topologies 
on the scaling relations in the MSQ, the Apollonian,
 and BA-like networks. 

\begin{figure}[htb] 
\includegraphics[height=90mm,angle=-90]{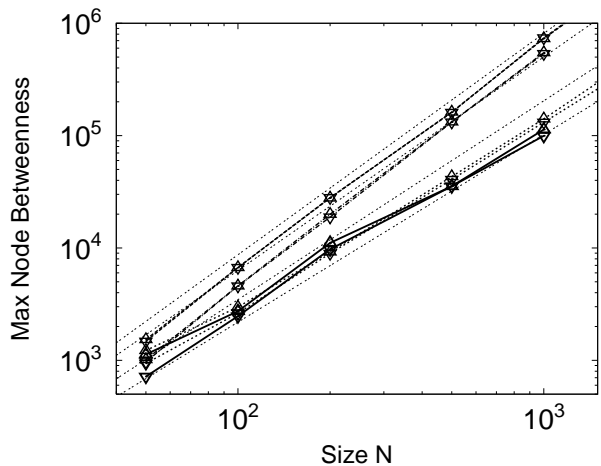} 
\caption{Scaling relations $N^{\delta}$ of the maximum node betweenness 
in the MSQ (solid line), BA-like:300 (dotted line),
the Apollonian (dashed-dotted line), and the 
BA-like:033 (dashed line) networks.
The triangle and inverted triangle marks 
correspond to Pop and Rand, respectively.
The thin lines without any marks 
guide the slopes $\delta = 1.67, 1.77, 1.87, 1.97$ 
from the bottom to the top.
} \label{fig_scaling_NodeBC}
\end{figure}

\begin{figure}[htb]
\includegraphics[height=90mm,angle=-90]{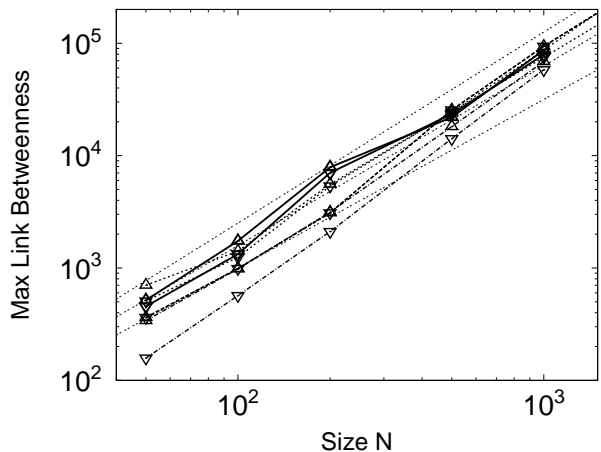} 
\caption{Scaling relations $N^{\delta}$ of the maximum link betweenness
in the MSQ, the BA-like, and the Apollonian networks. 
The thick lines and marks are the same as 
in Fig. \ref{fig_scaling_NodeBC}.
The thin lines without any marks 
guide the slopes $\delta = 1.5, 1.6, 1.7$ 
from the bottom to the top.
} \label{fig_scaling_LinkBC}
\end{figure}

Figure \ref{fig_scaling_NodeBC} shows 
the scaling relations of the maximum node betweenness.
They are separated into two groups of 
$\delta = 1.67 \sim 1.77$ and $\delta = 1.87 \sim 1.97$.
The MSQ network with only trimodal low degrees 
is located on the baseline.
The order of thick lines from the bottom to the top 
fairly corresponds to the increasing order of
largest degrees in these networks, 
as shown in Fig. \ref{fig_BA-like_Pk_LL}(a), 
because more packets tend to concentrate on large degree nodes as the
degrees are larger.
Figure \ref{fig_scaling_LinkBC} shows 
the scaling relations of the maximum link betweenness. 
The thick lines from the top to the bottom 
in the inverse order to that in Fig. \ref{fig_scaling_NodeBC} show that 
packets are distributed on many links connected to large degree nodes 
as the degrees are larger.
However, all lines lie almost 
around the intermediate slopes of $\delta = 1.6$, 
especially for a large size $N$, 
the maximum link load is at a similar level in the lowest case of the 
Apollonian network 
and the hightest case of the BA-like:300 or the MSQ network.
These results 
in Figs. \ref{fig_scaling_NodeBC} and \ref{fig_scaling_LinkBC}
are also consistent with the increasing orders of the maximum 
$B_{i}$ and $\bar{B}_{l}$ 
in Table \ref{table_correl_coef_BCs}.
Note that there exists no remarkable 
difference between the two patterns of 
Rand (marked by triangles)
and Pop (marked by inverted triangles) 
for both scaling relations of node and link loads 
except in the Apollonian network (dashed-dotted line).
In summary,
the MSQ network yields a better performance with a lower maximum load
than the other geographical networks, therefore 
it is more suitable for avoiding traffic congestion.

\begin{table}[htb]
\begin{center}
\begin{footnotesize}
\begin{tabular}{c|cc|cc} \hline
    & Rand        & Pop  & Rand        & Pop \\
Net & Max $B_{i}$ & Max $B_{i}$ 
& Max $\bar{B}_{l}$ & Max $\bar{B}_{l}$ \\ \hline
000 & 0.254 & 0.227 & 0.085 & 0.081 \\
003 & 0.766 & 0.854 & 0.048 & 0.087 \\
030 & 0.362 & 0.392 & 0.134 & 0.180 \\
{\bf 033} & {\bf 0.657} & {\bf 0.710} & {\bf 0.056} & {\bf 0.099} \\
{\bf 300} & {\bf 0.282} & {\bf 0.235} & {\bf 0.121} & {\bf 0.121} \\
303 & 0.444 & 0.532 & 0.073 & 0.107 \\
330 & 0.397 & 0.337 & 0.121 & 0.143 \\
333 & 0.634 & 0.620 & 0.106 & 0.072 \\ \hline
Apollonian & 0.295 & 0.278 & 0.056 & 0.059 \\ \hline
MSQ & 0.227 & 0.259 & 0.137 & 0.185 \\ \hline
\end{tabular}
\end{footnotesize}
\end{center}
\caption{Maximum betweenness centralities in 
the BA-like, the Apollonian, and the MSQ networks for $N = 100$. 
The triplet of 0 or 3 in the first column denotes the values of 
$\alpha \beta \gamma$ in the BA-like networks.
We note that the MSQ, the BA-like:300, the Apollonian, 
and the BA-like:033 networks are 
in increasing order of max $B_{i}$ 
and decreasing order of max $\bar{B}_{l}$.
} 
\label{table_correl_coef_BCs}
\end{table}

\section{CONCLUSION} \label{sec4}
Both mobile communication and wide-area wireless technologies
(or high-speed mass mobilities)
are important more and more to sustain our socio-economic activities,
however it is an issue to construct 
an efficient and robust network on the dynamic environment which 
depends on realistic communication (or transportation) requests.
In order to design the future ad hoc networks, 
we have considered geographical network constructions,
in which the spatial distribution of nodes is naturally determined 
according to a population in a self-organized manner.
In particular, the proposed MSQ network model is constructed by 
a self-similar tiling for load balancing of communication requests in
the territories of nodes.
On a combination of complex network science and computer science
approaches, this model has several advantages \cite{Hayashi09}: 
the robustness of connectivity, the bounded short paths, 
and the efficient decentralized routing on a planar network.
Moreover, we have numerically shown that 
the MSQ network is better in term of shorter 
link lengths than the geographical Apollonian and the BA-like networks 
which have various topologies ranging from river to SF networks.
As regards the traffic properties in the MSQ network,
the node load (defined by the maximum node betweenness) is lower 
with a smaller 
$\delta$ in the scaling relation $N^{\delta}$,
although the link load is at a similar level as that of 
the other geographical networks. 
Therefore,      
the MSQ network is more tolerant to traffic congestion
than the state-of-the-art geographical network models.

In a realistic situation,
packets are usually more often generated and received at a node
whose population is large in the territory.
Thus, we take into account spatially inhomogeneous
generations and removals of packets according to
a population. 
Concerning the effect of population on the traffic load, 
the heavy-loaded routes with much throughput of packets 
are observed between large population areas 
especially for a small size $N$.
The heavy-loaded parts are not trivial but localized 
depending on the geographical connection structure, 
whichever a spatial distribution of nodes or biased selections of the
source and the terminal is more dominant. 
In further studies, 
such spatially biased selections of packets 
should be investigated more to predict overloaded parts on a
realistic traffic, 
since the selections
may depend on other economic and social activities 
in tradings or community relationships 
beyond the activities related only to population, 
and probably affect the optimal topologies for the traffic 
\cite{Barthelemy06,Qian09,Nandi09}.
The optimal routing \cite{Sreenivasan07} instead of 
our shortest distance routing 
is also important for reducing the maximum betweenness 
($B_{i}$ or $\bar{B}_{l}$) and 
the exponent $\delta$ in the scaling relation.
In the optimizations that include 
other criteria \cite{Wang09,Gastner06b},
we will investigate the performance of the MSQ networks. 
Related discussions to 
urban street networks \cite{Cardillo06,Masucci09,Bitner09}
are attractive for investigating a common property which has 
arisen from the geographically pseudofractal structures.

\section*{Acknowledgment}
The authors would like to thank Prof. Tesuo Asano for suggesting the
proof in the Appendix
and  anonymous reviewers for their valuable comments.
This research is supported in part by a 
Grant-in-Aid  for Scientific Research in Japan, Grant No. 21500072.

\section*{Appendix}
We provide a sketch of the proof for the $t$-spanner property in the MSQ
networks. 
The following is similarly discussed 
for other polygons by exchanging the term 
``triangle'' or ``triangulation''
with other words (e.g., ``square'' or ``polygonal subdivision''), 
except that the maximum stretch factor $t$ may be greater than two 
for a general polygon. 
Remember that the self-similar tiling is obtained from the 
division by a contraction map as shown in Fig. \ref{fig_subdivision}.

Let ${\cal T}$ be a given triangulation and 
$L_{st}$ be a line segment
interconnecting two vertices $so$ and $te$ in ${\cal T}$. 
It intersects many triangle faces in ${\cal T}$.
Let $(so=v_0, v_1, \ldots, v_n=te)$ be those ordered vertices.
Now, we define a path using triangular edges.  For any two
consecutive vertices $v_i$ and $v_{i+1}$ there is a unique
triangle which contain both of them.  
In other words, $v_i$ and $v_{i+1}$ are the entrance and the exit 
of the line segment to the triangle.
For the pair $(v_i, v_{i+1})$
there are two paths, clockwise and anticlockwise, 
on the triangle.
We take the shorter one.  In this way, we can define a path on
the given triangulation.  If the path contains duplicated
segments with the opposite directions  
as shown in Fig. \ref{fig_typical_path}, 
we remove them.  Then, we have a path such that
each edge of the path is some triangular side.  
When the two consecutive edges on the path
belong to the links of a same triangle,
we take a shortcut by directly passing another 
edge (see two edges $v5'$-$v5''$ and $v5''$-$v6$ 
are replaced with $v5'$-$v6$ 
on the path in Fig.\ref{fig_typical_path}).
Since ${\cal T}$ is planar and there is no node inside each triangle, 
the length of this path is the closest to 
that of the line segment $L_{st}$,  
therefore the shortest in all other routes. 
As is easily seen,
in each triangle, the path length between $v_i$ and $v_{i+1}$ is
at most twice longer than the Euclidean distance between them.
The worst case of the maximum stretch factor 
$t=2$ is illustrated in Fig. \ref{fig_stretch}.


\end{document}